\documentclass[prb,a4paper,twocolumn,floatfix,showpacs,showkeys,amsmath,amssymb,nobibnotes,altaffilletter]{revtex4-1}

\usepackage{graphicx}
\usepackage{pslatex}
\usepackage{xspace}
\usepackage{color}
\usepackage{booktabs}

\newcommand{\etal}{\emph{et\,al.}\xspace}

\begin{document}

\title{Shockley Equation Parameters of P3HT:PCBM Solar Cells derived by Transient Techniques}

\author{A.~Foertig$^1$}
\author{J.~Rauh$^{1}$}
\author{V.~Dyakonov$^{1,2}$}\email{dyakonov@physik.uni-wuerzburg.de}
\author{C.~Deibel$^1$}\email{deibel@disorderedmatter.eu}
\affiliation{$^1$ Experimental Physics VI, Julius-Maximilians-University of W\"urzburg, D-97074 W\"urzburg, Germany}
\affiliation{$^2$ Bavarian Center for Applied Energy Research e.V. (ZAE Bayern), D-97074 W\"urzburg, Germany}

\date{\today}

\begin{abstract}
The Shockley equation (SE), originally derived to describe a p--n junction, was frequently used in the past to simulate current--voltage (j/V) characteristics of organic solar cells (OSC). In order to gain a more detailed understanding of recombination losses, we determined the SE parameters, i.e. the ideality factor and the dark saturation current, from temperature dependent static j/V-measurements on poly(3-hexylthiophene-2,5-diyl)(P3HT)\-:[6,6]-phenyl-C$_{61}$ butyric acid methyl ester (PCBM) bulk heterojunction solar cells. As we show here, these parameters are directly related to charge carrier recombination and become also accessible by transient photovoltage and photocurrent methods in the case of field-independent charge carrier generation. Although determined in very different ways, both SE parameters were found to be identical. The good agreement of static and transient approaches over a wide temperature range demonstrates the validity of the Shockley model for OSC based on material systems satisfying the requirement of field-independent polaron-pair dissociation. In particular, we were able to reproduce the photocurrent at various light intensities and temperatures from the respective nongeminate recombination rates. Furthermore, the temperature dependence of the dark saturation current $j_0$ allowed determining the effective band gap of the photoactive blend perfectly agreeing with the literature values of the energy onset of the photocurrent due to charge transfer absorption. We also present a consistent model directly relating the ideality factor to recombination of free with trapped charge carriers in an exponential density of tail states. We verify this finding by data from thermally stimulated current measurements.
\end{abstract}


\maketitle

\section{Introduction}
Organic solar cells (OSC) hold the potential to become a low-cost photovoltaic technology, produced in a roll-to-roll process and thus deserve serious consideration.\cite{li2012} Therefore, the efficiency of organic solar cells still has to be improved and a proper understanding of the elementary physical processes is required. Charge carrier recombination is one of the particularly crucial processes as it limits the electrical performance of OSC. Indeed, the polaron recombination dynamics fed into the continuity equation have been shown to deliver sufficient information to reproduce the complete j/V response of P3HT:PCBM solar cells at room temperature.\cite{shuttle2010}
Recently, Maurano \etal~\cite{maurano2010} successfully predicted the open circuit voltage of OSC based on four different polymer:fullerene compositions by studying nongeminate recombination. Thereby, authors combined parameters obtained by static (j/V) and transient photovoltage / transient photocurrent (TPV/TPC) methods in the Shockley equation.\cite{sze1981} In order to justify the validity of this approach and/or its limitations, we i) directly compared the ideality factor and dark saturation current density derived by the static and transient measurements and ii) reproduced the photocurrent for P3HT:PCBM OSC over the temperature range from 200 to 300 K at various light intensities. We demonstrate that identical Shockley parameters can indeed be determined from transient and static measurements making SE applicable to OSC, if the charge carrier generation is voltage independent. Deviations found at lower temperatures are attributed to voltage dependent charge carrier photogeneration due to the lack of thermal activation energy. Extending the work of Kirchartz \etal,\cite{kirchartz2011} we analyzed the impact of charge carrier recombination on the ideality factors in detail, and derived the dominant nongeminate loss mechanism.

\section{Theoretical Background}


The established model to describe the current--voltage response of semiconductor p--n junctions under illumination is based on the ideal SE,
\begin{align}	
	j(V)  & =  j_0\left(\exp \left( \frac{qV}{n_{id}kT}\right)-1\right)-j_{gen},
	\label{eqn:J_shockley}
\end{align}	
with dark saturation current $j_0$, elementary charge $q$, ideality factor $n_{id}$, thermal energy $kT$ and the photogenerated current $j_{gen}$.\cite{sze1981} Conventional p--n junctions are characterized by well developed energy-bands where photogeneration yields delocalized charge carriers. In contrast, OSC are based on excitonic materials, in which the photogeneration of quasi free polarons upon photon absorption is via bound precursor states and strongly depends on the active material. Furthermore, BHJ devices consist of donor and acceptor material phases across the whole volume, resulting in many spatially distributed heterojunctions instead of a single planar one.
 
Despite these differences to inorganic p--n junctions, the SE was successfully applied to organic solar cells based on different donor--acceptor systems in the past.\cite{chirvase2003,vandewal2009} Their application to devices based on MDMO-PPV:PCBM\cite{koster2005} failed due to voltage dependent photocurrent generation $j_{gen}(V)$. For OSC based on P3HT:PCBM, used in this work, the polaron pair dissociation is reported to be independent\cite{howard2010, kniepert2011, maurano2010} or weakly dependent\cite{limpinsel2010, deibel2009, mingebach2012} on voltage,  providing a good starting point for our analysis. 
For simplicity we assumed an ideal diode in this study, neglecting any series resistance ($R_s=0$) or leakages by shunts ($R_p=\infty$). No significant influence on the data was found by taking a series resistance into consideration, which is experimentally determined from the ohmic range of dark j/V-characteristics.
Eq.~(\ref{eqn:J_shockley}) solved for $V_{oc}$ at a given light intensity with $j(V_{oc})~=~0$ and for $j_{gen} \gg j_{0}$ results in
\begin{align}	
	V_{oc} &  =  n_{id} \frac{kT}{q} \ln \left( \frac{j_{gen}}{j_{0}} +1 \right) \\
	& \approx n_{id} \frac{kT}{q}\ln \frac{j_{gen}}{j_{0}}\cdot
	\label{eqn:Voc_shockley}
\end{align}	
We note that in order to experimentally access the photogenerated charges by j/V measurements, we assumed a voltage independent generation current $j_{gen}$.

\section{Experiment}
\subsection*{Sample preparation}
Bulk heterojunction (BHJ) solar cells were prepared by spin coating a 35~nm thick layer of poly(3,4-ethylendioxythiophene):polystyrolsulfonate (Clevios P VP AI 4083) on indium tin oxide samples with post-annealing step of 130$^{\circ}$C~for 10 minutes in a water-free environment. The P3HT:PCBM 1:0.8 blend made from solutions of 25~mg/ml in chlorobenzene was spin coated in an inert atmosphere resulting in 100--120~nm thick active layers. After annealing for 10~min at 130$^{\circ}$C~Ca (3~nm)/Al (100~nm) were evaporated thermally on top of the organic layer. P3HT was purchased from Rieke Metals (P200, $>$ 98~\% regioregular), PCBM (purity 99.5~\%) from Solenne. All materials were used without further purification.
Prior to any additional measurements an Oriel 1160 AM1.5G solar simulator was used to perform illuminated j/V--measurements of devices kept under inert glove-box atmosphere. 
The BHJ solar cells showed a power conversion efficiency of approximately 3.3~\%. The samples were transferred into a closed cycle optical cryostat for temperature dependent static and transient electrical studies.
\subsection*{Experimental techniques}
TPV measurements monitor the photovoltage decay upon a small optical perturbation during various constant bias light conditions. The voltage transient enable to determine the small perturbation charge carrier lifetime $\tau_{\Delta n}$ in dependence of the respective open circuit voltage due to the constant background illumination. TPC is measured with identical optical perturbation, but under short circuit conditions. Further experimental details can be found in Ref.~\onlinecite{shuttle2008},~\onlinecite{foertig2009}.
TPV/TPC measurements were performed using a 10 W white light LED (Seoul) and focusing optics for bias light illumination and an attenuated Nd:YAG laser flash ($\lambda=532$~nm, $<$ 80~ps pulse duration) to generate an additional small amount of charges within the device. The voltage transient was probed using a 1.5~G$\Omega$ impedance adapter to ensure open circuit conditions. Both, voltage and current  transients were acquired by a digital storage oscilloscope (Agilent Infinium DSO90254A). 

\section{Results and Discussion}

\begin{figure}
	\includegraphics[width=.9\linewidth]{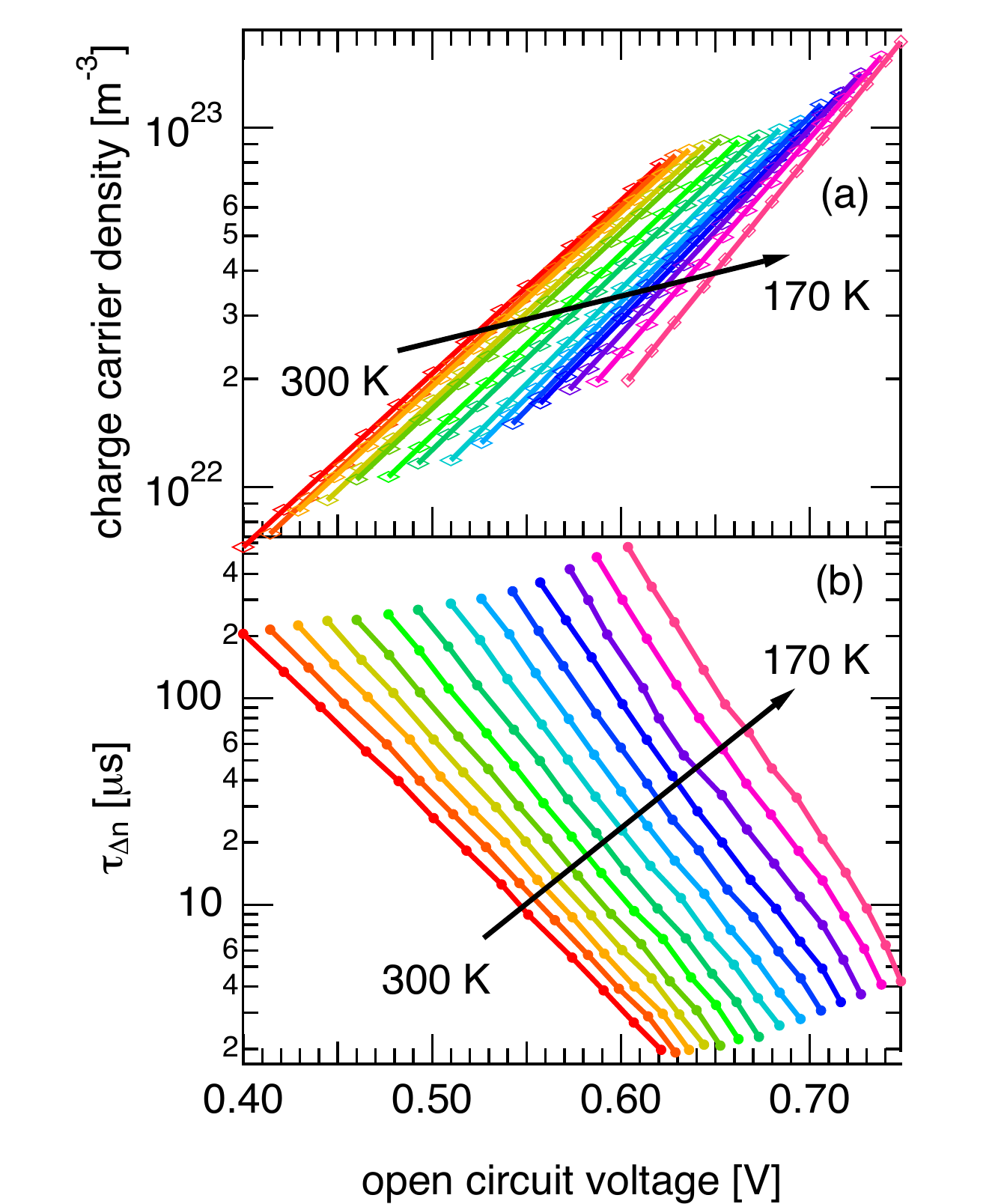}
	\caption{(Color Online) (a) Charge carrier density and (b) small perturbation charge carrier lifetime vs.\ open circuit voltage from TPV/TPC measurements for a solar cell based on P3HT:PCBM in dependence on temperature, as indicated by arrows. $V_{oc}$ was varied by changing the illumination intensity.}
	\label{fig:tau_n_voc}
\end{figure}

The results of temperature dependent TPV/TPC measurements are depicted in Fig.~\ref{fig:tau_n_voc}. As can be seen, charge carrier density $n$ (Fig.~\ref{fig:tau_n_voc} (a)) and small perturbation lifetime $\tau_{\Delta n}$ (Fig.~\ref{fig:tau_n_voc} (b)) both  exponentially depend on $V_{oc}$ and can be described by
\begin{align}
	n  & =  n_0 \exp \left( \frac{qV_{oc} }{n_n kT} \right) ,
	\label{eqn:n}
\end{align}	
with $n_n$ defined as the ideality factor of charge carrier density.
And similarly,
\begin{align}	
	\tau_{\Delta n} & = \tau_{\Delta n_0} \exp \left( -\frac{qV_{oc}}{n_\tau kT}  \right)
	\label{eqn:taun}	
\end{align}	
with the ideality factor of charge carrier lifetime $n_{\tau}$.
We note that the slopes in the semi-logarithmic representation of Fig.~\ref{fig:tau_n_voc} were previously described by parameters $\gamma$ and $\beta$,\cite{shuttle2008} which are related to our dimensionless ideality factors $n_n$ and $n_{\tau}$ by
\begin{align}
	n_n  & =  \frac{q}{\gamma kT},\\
	n_\tau & = \frac{q}{\beta kT} .
\end{align}
As seen in Fig.~\ref{fig:tau_n_voc} both slopes ($\beta ,\gamma$) increase when the temperature is lowered.

\subsection*{Equivalence of SE parameters from static and transient methods}

To describe the experimentally found polaron dynamics in organic photovoltaic devices a generalized equation $dn/dt=-k_\lambda n^{\lambda +1}=-R(n)$ is often used.\cite{shuttle2008, foertig2009, baumann2011} From TPV/TPC analysis,  the decay order of the recombination rate $R$,  $\lambda +1$, and the small perturbation charge carrier lifetime $\tau_{\Delta n}$ can be experimentally determined. Using these values the total charge carrier lifetime $\tau_n$ can be calculated, as it was shown in Ref.~\onlinecite{Hamilton2010},
\begin{equation}
	\tau_n  =  \tau_{\Delta n}(\lambda +1),
	 \label{eqn:lambda}	
\end{equation}
where $\lambda  =  \frac{n_n}{n_\tau}$. Using Eqs.~(\ref{eqn:taun}) and (\ref{eqn:lambda}), small perturbation lifetimes values can be eliminated, 
\begin{align}
	\tau_n & = \tau_{n_0}\exp \left(  -\frac{qV_{oc}}{n_\tau kT}   \right) 
	\label{eqn:tau} .
\end{align}
Dividing Eq.~(\ref{eqn:n}) by Eq.~(\ref{eqn:tau}), we obtain:
\begin{align}
	\underbrace{\frac{n}{\tau_{n}}}_{R(n)} & = \underbrace{\frac{n_0}{\tau_{n_0}}}_{R_0} \exp \left( \frac{qV_{oc}}{(\frac{1}{n_n}+\frac{1}{n_\tau})kT}\right) 
	\label{eqn:ntau} 
\end{align}
where $R(n)$ is defined as the recombination rate under illumination, depending on illumination intensity and $R_0$ the recombination rate in dark. Solving Eq.~(\ref{eqn:ntau}) for $V_{oc}$ yields a more generalized expression,
\begin{align}
	V_{oc} & = n_R \frac{kT}{q} \ln \left( \frac{R(n)}{R_0}  \right)  
	\label{eqn:Voc_tpv},
\end{align}
with the recombination ideality factor $n_R$ defined as
\begin{align}
           n_R^{-1} & = n_n^{-1}+n_{\tau}^{-1} ,
           \label{eqn:n_R}
\end{align}
which is consistent with earlier representations of the ideality factor.\cite{vanberkel1993,kirchartz2011}

Comparing Eq.~(\ref{eqn:Voc_tpv}), derived by using relations empirically found in transient measurements (Eqs.~(\ref{eqn:n}),(\ref{eqn:taun})), and the right hand side of Eq.~(\ref{eqn:Voc_shockley}), the SE solved for $V_{oc}$ to describe steady-state j/V-measurements, the same equation structure becomes apparent. To compare both equations, first the equality of recombination rates and respective currents is motivated.
At $V_{oc}$ generation and recombination rates are equal, i.e. $G=R$, which implies the generation current $j_{gen}$ (see Eq.~(\ref{eqn:Voc_shockley})) is cancelled by the recombination current $j_{loss}$, defined as 
\begin{align}
	j_{loss} & = R(n)qd
	\label{eqn:jloss},
\end{align}
with the thickness of the active layer $d$. 
Likewise, the dark saturation current $j_0$ may be treated as thermally generated intrinsic charge carriers with the density $n_0$ and lifetime $\tau_{\Delta n_0}$ canceled by recombination in thermal equilibrium. Hence, the respective loss current in the dark $j_{loss_0}$ can be written as
\begin{align}
	j_{loss_0} & = R_0qd
	\label{eqn:jloss0} .
\end{align}

\begin{figure}
	\includegraphics[width=.9\linewidth]{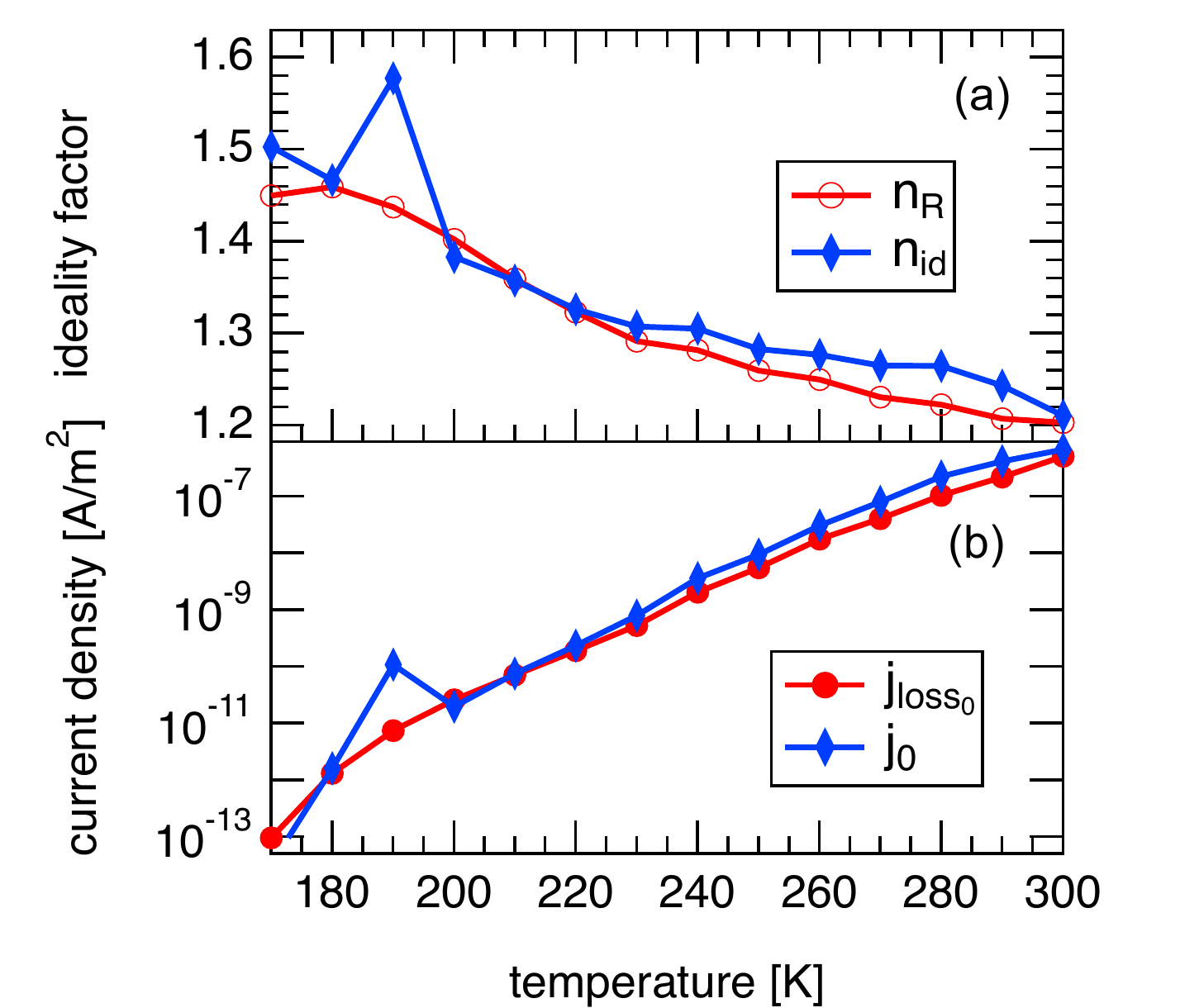}
	\caption{(Color Online) (a) Ideality factor of a P3HT:PCBM 1:0.8 solar cell in dependence on temperature revealed from TPV/TPC studies on the recombination dynamics $n_R$ and from illuminated j/V data $n_{id}$, analyzed with the respective Shockley equation~(see Eq.~(\ref{eqn:Voc_shockley})). (b) Respective dark saturation current densities in dependence on temperature. $j_{loss_0}$ was calculated from TPV/TPC studies on the recombination dynamics in the dark with Eq.~(\ref{eqn:jloss0}) and $j_{0}$ from illuminated j/V response at $V_{oc}$ via Eq.~(\ref{eqn:Voc_shockley}).}
	\label{fig:comp}
\end{figure}

The above considerations provide a rationale to the interconnection between the SE parameters derived diversely. 
Fig.~\ref{fig:comp}~(a) shows ideality factors $n_{id}$ and $n_R$, as defined above, which were experimentally obtained as function of temperature by means of static (j/V) and transient (TPV/TPC) methods, respectively. To determine the static SE parameters j/V characteristics were recorded for different bias lights. Then, $V_{oc}$ was plotted vs.\ $j_{gen}$ and analyzed by Eq.~(\ref{eqn:Voc_shockley}) to yield the ideality factor $n_{id}$. Within the measured temperature range $n_R$ and $n_{id}$ correspond well to each other. 
Similary, the dark saturation current $j_0$ and the dark recombination current $j_{loss_0}$, depicted in Fig.~\ref{fig:comp}~(b), are almost equal for different temperatures. The ideality factor rises from 1.2 at room temperature to 1.5 at 170 K. A similar trend with temperature was observed  in the past for organic solar cells based on different materials, although a stronger temperature dependence was obtained.\cite{koster2005, riedel2005}

From the experimental observation shown in Fig.~\ref{fig:comp}, the SE equivalence for P3HT:PCBM based OSC to Eq.~(\ref{eqn:Voc_tpv}), derived from studying charge carrier recombination by transient techniques, can be suggested.

\subsection*{Effective band gap energy $E_g$}

\begin{figure}
	\includegraphics[width=.9\linewidth]{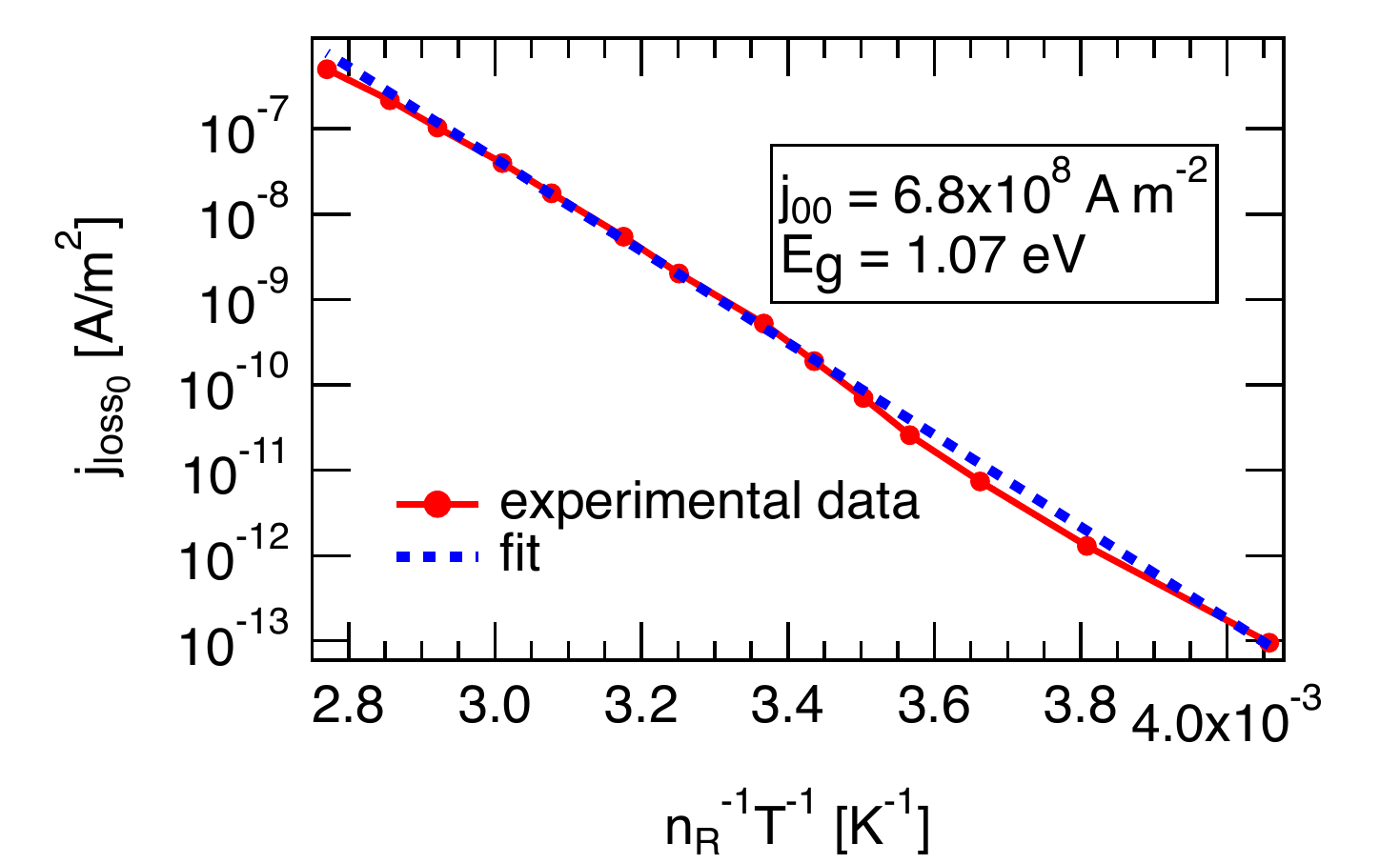}
	\caption{(Color Online) Dark saturation current evaluated by studying the charge carrier dynamics plotted versus $1/Tn(T)$ in order to estimate the effective band gap energy $E_g$ from Eq.~(\ref{eqn:j_0}).}
	\label{fig:j0_invnT}
\end{figure}

We first focus on the strong temperature dependence of the dark saturation current (Fig.~\ref{fig:comp}~(b)). In Fig.~\ref{fig:j0_invnT}, the dark saturation current $j_{loss_0}$ determined by TPV/TPC is plotted versus $1/(Tn_R(T))$, with the temperature dependent ideality factor $n_R(T)$ from Fig.~\ref{fig:comp}~(a). The linear fit of the semi-logarithmic plot is in excellent agreement with the measured data and is expressed by the relation
\begin{align}
	j_{loss_0} & =  j_{00}\exp \left(-\frac{E_g}{n_R(T)kT}\right) .
	\label{eqn:j_0} 
\end{align}
The fit demonstrates that the dark saturation current follows the Boltzmann distribution and allows to identify the effective band gap energy $E_g$, which is proportional to the difference between HOMO and LUMO energy levels of donor and acceptor, respectively. We found $E_g=~1.07~$eV for P3HT:PCBM. From the $V_{oc} (T)$ data (not shown), a value of $E_g\approx 0.9$~eV was evaluated according to Ref.~\onlinecite{rauh2011}. Vandewal et.~al  obtained the value of $E_g=1.08$~eV from the energy onset of the photocurrent generated by CT absorption,\cite{vandewal2008} which agrees particularly well with our value. Guan et  al.\ found somewhat larger values ($E_g\approx1.3-1.4~$eV) using ultraviolet and inverse photoemission spectroscopy.\cite{guan2010} As discussed in Ref.~\onlinecite{blakesley2011}, energy values derived by photoemission spectroscopy can be seen as onsets of respective valence and conduction band. Due to charge carrier relaxation into the tail states of the density of states, the techniques such as TPV/TPC and j/V-measurements are expected to reveal the effective bandgap $E_g$ from the relaxed density of occupied states instead.

\subsection*{Loss current vs.\ photocurrent}

To complete the comparison of Eq.~(\ref{eqn:Voc_shockley}) and Eq.~(\ref{eqn:Voc_tpv}) we consider the photogenerated charge carrier density, being equal to the density of charges recombining at $V_{oc}$. In order to estimate the photogenerated charge carriers, we used the saturated current density $j_{sat}$ under illumination in reverse bias ($V=-1~\text{V}$).

If the generation of charge carriers is voltage independent for P3HT:PCBM cells in the range of $-1~\text{V}<V<V_{oc}$\cite{kniepert2011, mingebach2012} and $j_{sat}$ is not reduced by nongeminate recombination, we expect
\begin{align}
	j_{sat}  & = j_{loss}
	\label{eqn:jloss_comp} .
\end{align}
For comparison, Eq.~(\ref{eqn:jloss}) was used to calculate the recombination current $j_{loss}$ under illumination at open circuit.

\begin{figure}
	\includegraphics[width=.9\linewidth]{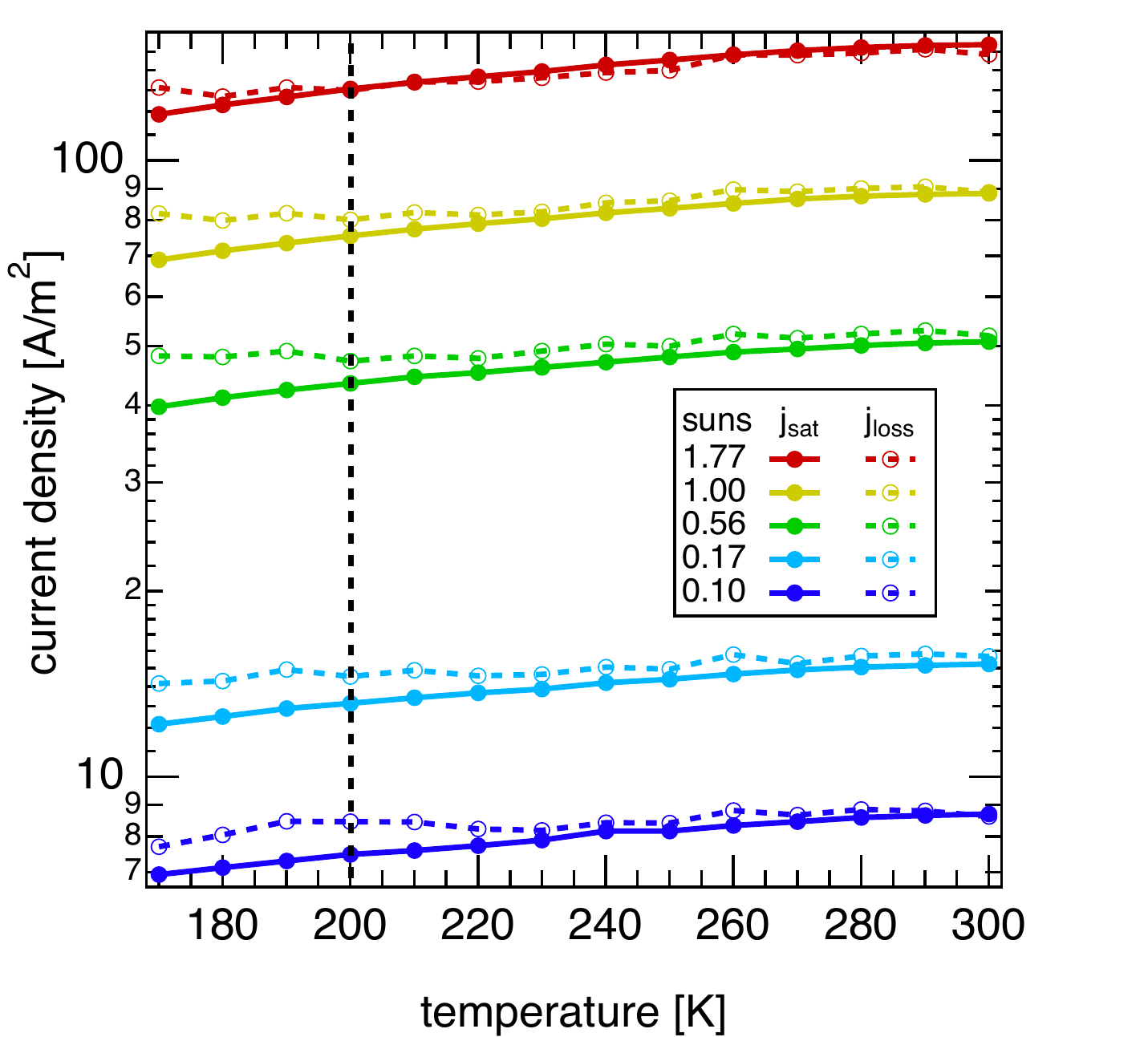}
	\caption{(Color Online) Current density $j_{sat}$ (solid lines) from j/V response compared to recombination current $j_{loss}$ (dashed lines) measured by TPV/TPC for different temperatures and bias lights according to Eq.~(\ref{eqn:jloss_comp}). For $T<$200~K a stronger deviation of both parameters becomes apparent, indicated by the dashed vertical line.}
	\label{fig:jloss_T}
\end{figure}

The saturation current $j_{sat}$ and the nongeminate recombination current $j_{loss}$ at $V_{oc}$ determined via TPV/TPC measurements for different illumination intensities from 0.1 sun to about 1.8 suns are shown in Fig.~\ref{fig:jloss_T}. Within experimental error, $j_{loss}$ is in a very good agreement with $j_{sat}$ for all light intensities and temperatures above 200~K.
Below this temperature, $j_{sat}$ becomes smaller than $j_{loss}$, which seems to imply more charges recombining than being generated. 
This counterintuitive result can at least partly be explained by estimating the nongeminate recombination losses at $V=-1~\text{V}$ according to Ref.~\onlinecite{koster2011}. We find 2\% loss at room temperature and 5\% at 170~K, revealing the limitations of the analysis as the generation current is underestimated. Accounting for the nongeminate losses, a very small deviation of $j_{sat}$ and $j_{loss}$ remains, showing that the field dependence of the photogeneration is less than 10\% between open and short circuit conditions even at low temperatures. In accordance with earlier findings,\cite{limpinsel2010, deibel2009, kniepert2011, mingebach2012} the photocurrent in P3HT:PCBM solar cells is at best slightly field dependent in this voltage range.

\subsection*{Contributions to recombination ideality factor $n_R$}

\begin{figure}
	\includegraphics[width=.9\linewidth]{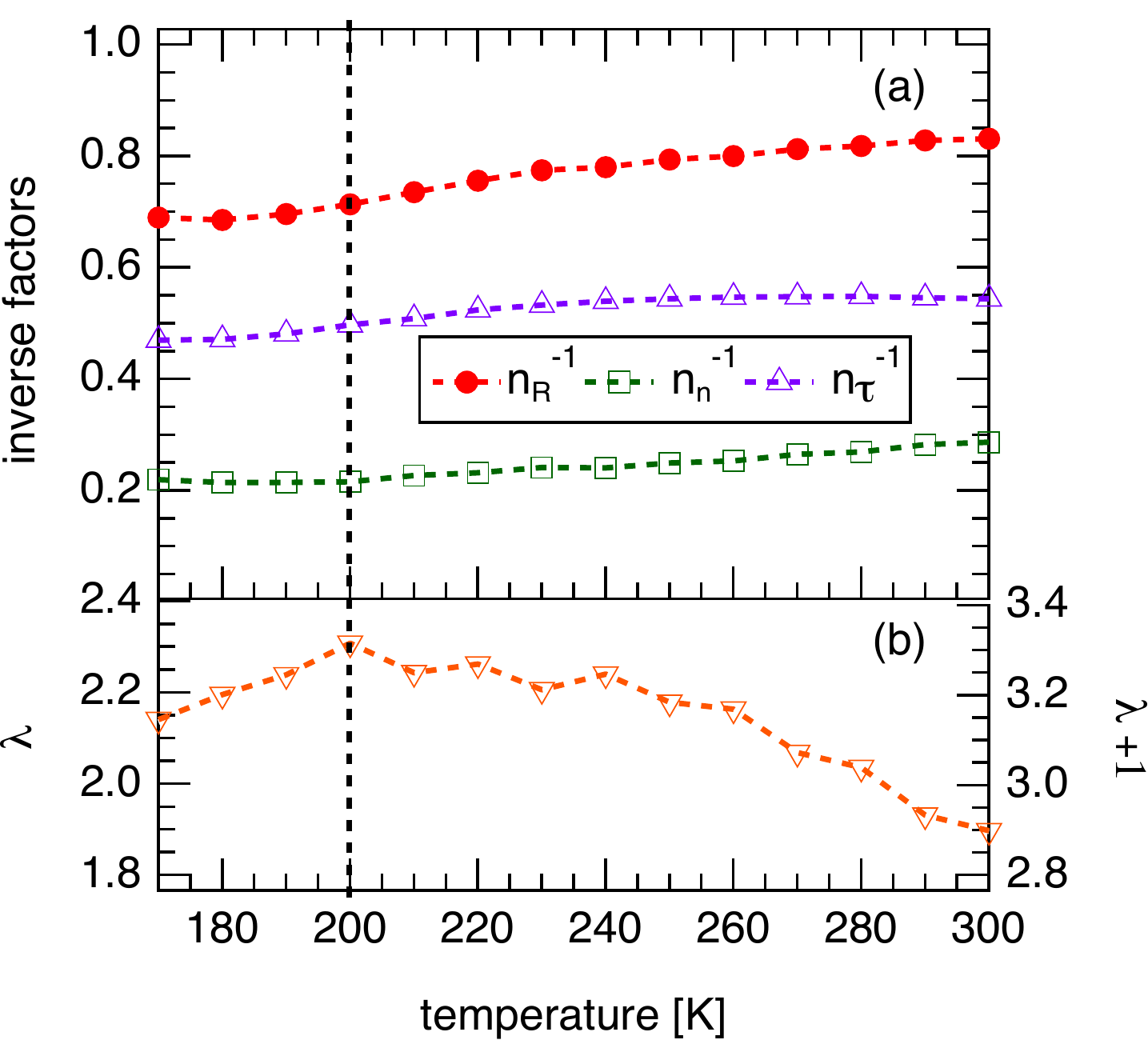}
	\caption{(Color Online) (a) Temperature dependence of the inverse ideality factors for  charge carrier density, $n_n^{-1}$, and lifetime, $n_{\tau}^{-1}$, as well as their sum, resulting in the inverse ideality factor $n_R^{-1}$. (b) Corresponding recombination parameter $\lambda=n_n/n_\tau$ (left) and recombination order $\lambda+1$ (right).}
	\label{fig:nr}
\end{figure}

In Eq.~(\ref{eqn:Voc_tpv}) the recombination ideality factor $n_R$ was defined. As mentioned above $n_R$ is composed by the ideality factor for charge carrier density $n_n$ and charge carrier lifetime $n_{\tau}$, respectively (see Eq.~(\ref{eqn:n_R})). In Fig.~\ref{fig:nr}~(a),  the inverse ideality factors derived from the transient experiments are plotted versus temperature. At 300~K, we found $n_\tau=1.84$ and $n_n=3.49$, yielding a recombination ideality factor of $n_R=1.20$. As shown already in Fig.~\ref{fig:comp}~(a), the ideality factor from the j/V-characteristics under illumination was $n_{id}=1.21$. Although we focussed on the ideality factors measured under illumination, we state for comparison that the dark ideality factor was 1.30, whereas Kirchartz \etal\cite{kirchartz2011} found a value of 1.6. 

The recombination order, given by $\lambda+1=\frac{n_n}{n_\tau}+1$ as pointed out above, is shown in Fig.~\ref{fig:nr}~(b). The recombination order at 300~K of $\lambda+1=2.90$ was slightly higher as compared to the values of 2.6 (Ref.~\onlinecite{shuttle2008}) or 2.75 (Ref.~\onlinecite{foertig2009}).

\subsection*{Interpretation of ideality factors: Recombination mechanism}

Based on the very concise approach offered by Kirchartz \etal,\cite{kirchartz2011} we used the experimentally determined ideality factors to consider the dominant recombination mechanism in the P3HT:PCBM solar cells under open circuit conditions. 

Ideality factors higher than 1 in organic bulk heterojunction solar cells are commonly seen as an evidence for a dominant trap assisted recombination process. Kirchartz \etal\ proposed the recombination of free charge carriers $n_c$ with trapped charge carriers $n_t$ in an exponential tail as dominant nongeminate loss mechanism. The corresponding recombination rate is $R \propto n_c n_t$, assuming symmetric electron and hole concentrations. Using the relation, $n_t \propto n_c^{kT/E_U}$,\cite{adriaenssens1995} of free $n_c$ to trapped $n_t$ charge carriers they were able to calculate the characteristic (Urbach) tail energy $E_U$ from experimental data\cite{shuttle2008} by three routes: (i) By the voltage dependence of the extracted charge carrier concentration, which---rewritten in our notation---is given by the ideality factor $n_n$,
\begin{align}
	E_U & = \frac{n_n kT}{2} ,
	\label{eqn:Eu_nn} 
\end{align}
using their experimental $n_n\approx 4.2$.  (ii) From the (in their case dark) ideality factor, based on the relation by van Berkel \etal,\cite{vanberkel1993} 
\begin{align}
	n_{id}^{-1} & = \frac{kT}{2E_U} + \frac{1}{2},
	\label{eqn:vanBerkel} 
\end{align}
which corresponds to our Eq.~(\ref{eqn:n_R}), assuming $n_\tau=2$ and replacing $n_n$ by using Eq.~(\ref{eqn:Eu_nn}). (iii) From the recombination order 
\begin{align}
	\lambda+1=\frac{E_U}{kT}+1 , 
	\label{eqn:lambda_Kirchartz} 
\end{align}
with an experimentally determined recombination order of 2.6.\cite{shuttle2008} From (i) Kirchartz \etal\ found $E_U\approx 50$~meV, from (ii) $E_U\approx 100$~meV at room temperature, whereas from (iii) $E_U=41$~meV can be calculated. This discrepancy was discussed in some detail in their publication.\cite{kirchartz2011}

We considered a more general recombination rate, including the annihilation of free charge carriers with one another and free with trapped charge carriers, 
\begin{align}
	R(n)=k' n_c (n_c+n_t) = k' n_c n .
	\label{eqn:R}
\end{align}
Here, the overall charge carrier concentration is given as $n=n_c+n_t$. For simplicity, we used the same recombination prefactor $k'$ for both contributions $n_c^2$ and $n_c n_t$. 

Usually the rate limiting step in nongeminate recombination is the finding of the actually localized charge carriers by a hopping process, which is reflected by a prefactor proportional to the mobility of the mobile charge carriers, based on Langevin theory.\cite{deibel2010review,kirchartz2011,wetzelaer2011b}
However, although beyond the scope of this article, we point out that in the multiple-trapping-and-release approximation of hopping transport, the effective mobility $\mu$ is proportional to a trap-free mobility $ \mu_0$ times the fraction of free to all charge carriers, $\mu_0\frac{n_c}{n}$. Within this approach, $R\propto \mu n^2$ and $R\propto \mu_0 n_c n$ are in principle equivalent. We ask the reader to bear in mind that this representation can serve as a first approximation, but neglects the field dependence in a hopping model. For our purpose here, however, Eq.~(\ref{eqn:R}) is sufficient to describe the recombination process in terms of the ideality factors discussed above.

The small-signal method TPV yields the effective lifetime of all charge carriers, $\tau_n$. For charge extraction, we assumed that all charge carriers were extracted, in accordance with Kirchartz \etal\cite{kirchartz2011} Thus, our experimentally determined effective recombination rate is $R=n/\tau_n=(n_c+n_t)/\tau_n$ (c.f.\ Eq.~(\ref{eqn:ntau})). Comparing our assumption for the  recombination rate, Eq.~(\ref{eqn:R}) to our effective recombination rate, we find 
\begin{align}
	\tau_n = (k' n_c)^{-1} .
	\label{eqn:tau_for_R}
\end{align}

In order to test our assumption, we considered the voltage dependence of the constituents $n$ and $\tau_n$ of the effective recombination rate by investigating their ideality factors $n_n$ and $n_\tau$, respectively. Using the exponential tail model,\cite{kirchartz2011} we also followed the three different routes outlined above. Calculating the characteristic tail energy (i) directly from $n_n$, we determined $E_U=45$~meV at room temperature. By route (ii), using our Eq.~(\ref{eqn:n_R}) instead of Eq.~(\ref{eqn:vanBerkel}), we again found 45~meV with either $n_R$ or (illuminated) $n_{id}$. In contrast to Kirchartz \etal,\cite{kirchartz2011}, our result from (ii) is consistent with (i). We point out that the authors had a rather high ideality factor of 1.6 which was determined under dark conditions, in contrast to our consistent values of $n_{id}\approx n_R=1.20$. When using our dark ideality factor of 1.30, we find 48~meV for the characteristic tail energy. Deriving the tail energy from the recombination order, route~(iii),
\begin{align}
	\lambda+1 & = \frac{n_n}{n_\tau}+1\\
	& = \frac{E_U}{kT}\frac{2}{n_\tau}+1 , 
\end{align}
using Eq.~(\ref{eqn:Eu_nn}), yielded again a result compatible with the other two approaches, as essentially always the same equations are used for all routes within our consistent framework. In contrast to Eq.~(\ref{eqn:vanBerkel}), as used in Ref.~\onlinecite{kirchartz2011}, deviations of $n_\tau$ from the value of 2 are considered. We point out that due to our focus on the interpretation of ideality factors here, which is only qualitative in view of the recombination rates, we are not able to make statements concerning the contributions from delayed recombination.\cite{baumann2011,rauh2012} For understanding the consistency of the charge carrier concentration and voltage dependence of the loss current in view of recombination order and ideality factor, see Appendix~\ref{appendix_jloss}.

The increase of $n_n$ with lower temperatures (Fig.~\ref{fig:nr}~(a)) also indicates a thermally activated process concerning the charge carrier concentration: the lack of thermal energy leading to a growing fraction of trapped charge carriers is compatible with the assumption of exponential tail states. Calculating the characteristic Urbach energy from the temperature dependent $n_n$  by using Eq.~(\ref{eqn:Eu_nn}), we found $E_U=40$~meV for 200~K, showing that an exponential tail may not be the precise shape of the density of trap states, but can serve as an approximation in a limited temperature range.

Within our assumption of the recombination rate, Eq.~(\ref{eqn:R}), the effective lifetime is inversely proportional to the free charge carrier concentration $n_c$ (Eq.~(\ref{eqn:tau_for_R})). Therefore, we would expect that $n_\tau$ should equal 2 in accordance with $n_n$ for free charge carriers. However, this statement is only valid if the recombination prefactor $k'$ is assumed to be voltage independent. While we cannot determine the detailed reason for our experimental finding of $n_\tau=1.84$, we point out that the assumption of $k'\not = k'(V)$ may not hold true\cite{shuttle2010b,rauh2012}  and that concentration gradients are disregarded.

\begin{figure}
	\includegraphics[width=.9\linewidth]{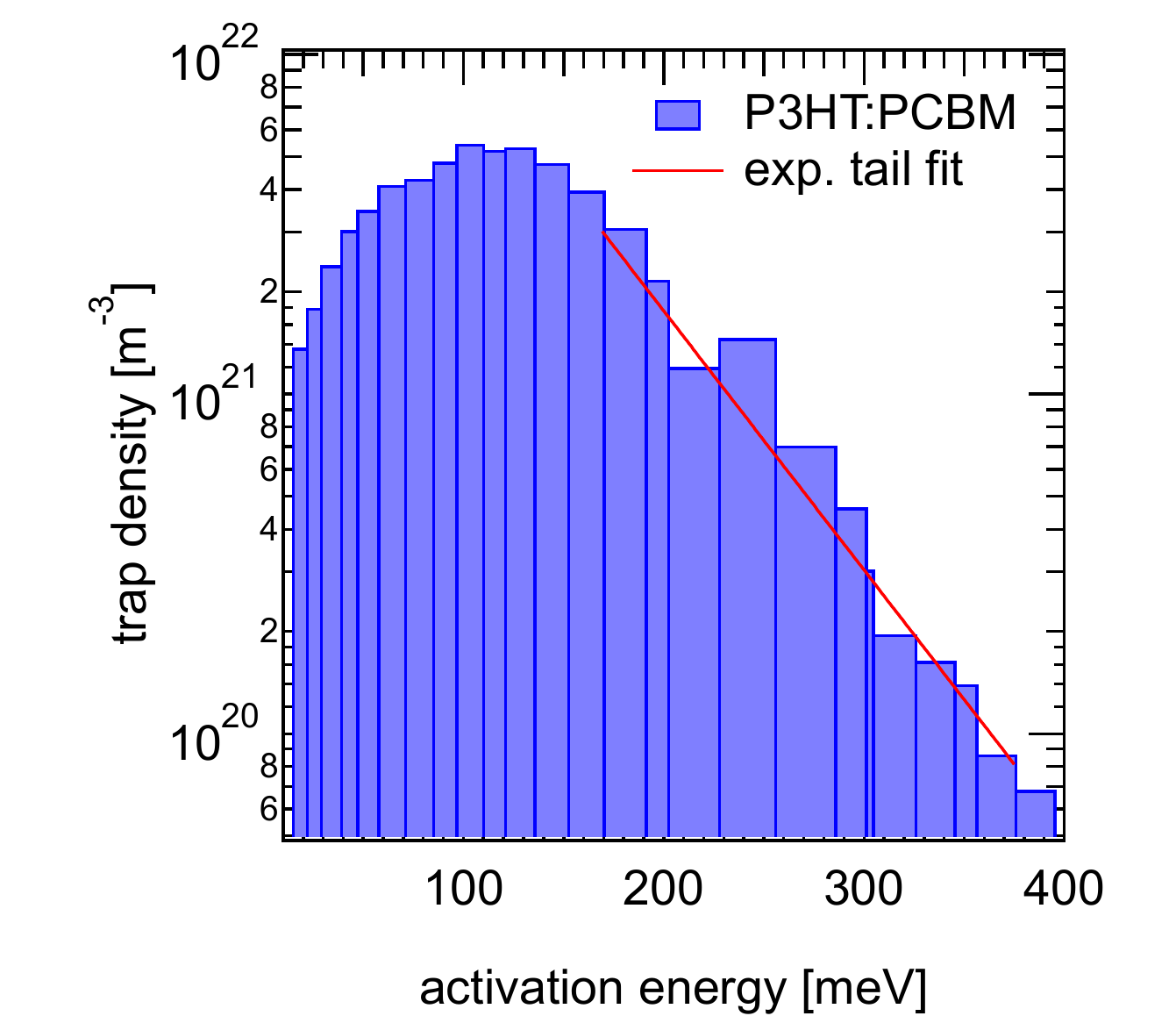}
	\caption{(Color Online) Trap density distribution of a P3HT:PCBM blend device, as measured by the thermally stimulated currents technique.\cite{schafferhans2010} Disregarding the detailed shape of the trap distribution, we approximated the energetic tail (solid line), finding a characteristic energy $E_U \approx 57$~meV.}
	\label{fig:exp-tail}
\end{figure}

In order to verify that the nongeminate recombination mechanism in annealed P3HT:PCBM solar cells is indeed due to losses of free carriers with one another and with carriers trapped in exponential tail states with $E_U\approx 50$~meV, we reconsidered thermally stimulated current measurements presented previously.\cite{schafferhans2010} In this set of experiments, we did not account for trap states deeper than 400~meV. The distribution of the trap states is shown in a semilogarithmic plot (Fig.~\ref{fig:exp-tail}). Neglecting the detailed shape, we approximated the energetic tail by an exponential Urbach fit. Our analysis yielded a characteristic energy $E_U\approx 57$~meV, in accordance with a recent reconstruction of the density of trap states in P3HT:PCBM blends,\cite{mackenzie2012} and verifies our findings from above.

\section{Conclusion}

The SE parameters such as ideality factor $n_{id}$ and dark saturation current $j_0$ were derived either from static j/V-measurements or from the transient techniques (TPV/TPC), were shown to coincide within experimental error for OSC based on P3HT:PCBM. We directly compared the ideality factor and dark saturation current density experimentally and reproduced the saturation photocurrent for P3HT:PCBM OSC over the temperature range from 200 to 300 K at various light intensities. Recombination current determined under open circuit conditions is shown to be equal to the respective saturation current, which implies that the polaron pair dissociation is not significantly influenced by the electric field. A good agreement of static and transient approaches over a wide temperature range demonstrates the validity of the Shockley model for OSC, if the charge carrier photogeneration is voltage independent.  Additionally, the dark saturation current measured at different temperatures was used to determine the effective band gap of P3HT:PCBM blend to be in the range of $E_g\approx0.9-1.1$~eV, which is in good agreement to values from literature. Considering the ideality factors, we found that nongeminate recombination of free with both free and trapped charge carriers in tail states is the dominant loss mechanism. Using data from thermally stimulated current measurements, we verified that the charge carrier traps can indeed be approximated by an exponential trap distribution.

\begin{acknowledgments}
The authors would like to thank T.~Kirchartz (Imperial College, London) for helpful inspiration to parts of this work as well as D.~Rauh and J.~Lorrmann for discussions and reading the manuscript. 
A.F.'s work was financed by the European Commission in the framework of the Dephotex Project (Grant No.~214459). C.D. gratefully acknowledges the support of the Bavarian Academy of Sciences and Humanities. V.D.'s work at the ZAE Bayern is financed by the Bavarian Ministry of Economic Affairs, Infrastructure, Transport and Technology.
\end{acknowledgments}

\appendix*

\section{Voltage and charge carrier concentration dependence of the loss current}\label{appendix_jloss}

In Eq.~(16) of Ref.~\onlinecite{kirchartz2011}, Kirchartz \etal\ considered the carrier concentration dependent loss current at $V_{oc}$,
\begin{align}
	j_{loss}  & \propto R(n) \propto n^\delta \equiv n^{\lambda+1}
	\label{eqn:jloss_n} .
\end{align}
and its connection to the voltage dependence of the carrier concentration, c.f.\ our Eq.~(\ref{eqn:n}). 

They found that
\begin{align}
	\underbrace{\frac{d\ln j_{loss}}{d\ln n}}_{\lambda+1}  & \cdot \underbrace{\frac{d\ln n}{dV}}_{n_n^{-1} \frac{q}{kT}} = \underbrace{\frac{d\ln j_{loss}}{dV}}_{n_{id}^{-1} \frac{q}{kT}}
	\label{eqn:consistency1} 
\end{align}
has to hold for self consistency (in our notation). Evaluating this equation, considering $\lambda=\frac{n_n}{n_\tau}$, we see that
\begin{align}
	\left(\frac{n_n}{n_\tau}+1\right)  & \cdot n_n^{-1} = n_{id}^{-1}
	\label{eqn:consistency2} 
\end{align}
directly yields Eq.~(\ref{eqn:n_R}).

\end{document}